# Highly (111)-orientated BiFeO$_3$ thin film deposited on La$_{0.67}$Sr$_{0.33}$MnO$_3$ buffered Pt/TiO$_2$/SiO$_2$/Si (100) substrate

Qingqing Ke, Wenlai Lu, Xuelian Huang, and John Wang[a]

*Department of Materials Science and Engineering, National University of Singapore, Singapore 117574*

## Abstract

Multiferroic BiFeO$_3$ (BFO) thin film exhibiting desired ferroelectric and enhanced magnetic properties was grown on La$_{0.67}$Sr$_{0.33}$MnO$_3$ (LSMO) buffered Pt/TiO$_2$/SiO$_2$/Si substrates by off-axis RF magnetic sputtering, where a highly (111)-oriented texture was obtained. The BFO/LSMO thin film exhibits excellent ferroelectric and dielectric behaviors (2$Pr$ ~210.7 μC/cm$^2$, 2$Ec$~435 kV/cm, $\varepsilon_r$ ~116.8, and tanδ ~ 2.7% at 1 kHz), together with a long fatigue endurance up to 10$^{10}$ switching cycles at amplitude of 300 kV/cm. An enhancement in magnetic behavior was also observed with $Ms$=89.5 emu/cm$^3$, which is largely contributed from the magnetic layer of LSMO. The coexistence of ferroelectric and ferromagnetic properties in the double layered BFO/LSMO thin film makes it a promising candidate system for applications where the magnetoelectric behavior is required.

**Keywords**: BiFeO$_3$; LSMO; (111) Orientation; Ferroelectric properties

PACS codes: 77.22.Gm, 77.55.-g, 77.55.Nv

---

a) Author to whom corresponding should be addressed. Electronic mail: msewangj@nus.edu.sg

Introduction

BiFeO$_3$ (BFO), as one of the very few known multiferroic materials at room temperature, has drawn considerable interest owing to its high ferroelectric ($T_c$~1103 K) and antiferromagnetic ($T_N$~643 K) transition temperatures and therefore potential applications in a wide spectrum of spintronics, data storage, and microelectromechanical devices.[1-3] Its ferroelectric and magnetic behaviors are closely dependent on the orientation of the produced BFO film, which can be modulated by using different substrates and processing conditions. For example, the substrates possessing similar lattice constant as BFO, such as SrTiO$_3$ (STO)[4] or LaAlO$_3$ (LAO),[5] have been used to deposit BFO thin films with controlled orientation. A polarization of 50~80 μC/cm$^2$ has been observed with (001) and (110) orientations,[6] while a much higher polarization value of $2Pr$=200 μC/cm$^2$ has been measured with the BFO thin film of (111) orientation, where the spontaneous polarization lies in the (111) direction.[7] Moreover, it is worthy mentioning that the magnetic structure of BFO thin film is also related to the substrate orientation which can introduce the different strain–relaxation processes and gives rise to different thin-film crystal structures.[8] However, the use of single crystals as substrates may constrain the applications of BFO films, as they are of high cost and incompatible with the well-established Si technology. To use Si wafers as substrates for BFO thin films, an appropriate buffer layer, such as SrRuO$_3$ (SRO)[9] and LaNiO$_3$ (LNO),[10] has been inserted between BFO and Pt coated Si substrates to improve the growth of BFO thin film and control its orientation in certain cases.[9,10] Compared to the buffer layers of SRO and LNO, LSMO is an attractive candidate as a bottom electrode material for ferroelectric perovskite oxide films, considering its high Curie temperature of 390 K and good electrical conductivity.[11] Indeed, several intriguing ferroelectric and magnetoelectric behaviors have been observed with the BFO/LSMO heterostructures.[12] However, the research work on the polycrystalline of BFO on the LSMO buffer layer is limited and the synthesized films always show poor electrical and polarization behaviors. In addition, to the best of our knowledge, there are few reports on the success of developing a highly (111)-oriented BFO thin film buffered with LSMO on the platinized Si substrates with the desired ferroelectric behavior, such as a large remnant polarization, long ferroelectric fatigue resistance and low leakage current.

In the present work, the BFO thin film was deposited on the LSMO buffered Pt/TiO$_2$/SiO$_2$/Si substrates by off-axis radio frequency magnetic sputtering, where a highly (111)-oriented texture was developed. The impact of LSMO buffer layer on the electric and magnetic properties of BFO thin films is investigated. The highly (111)-oriented texture, arising from the LSMO buffer layer, gives rise to an apparent improvement in ferroelectric behavior, with a high 2*Pr* value of 210.7 μC/cm$^2$ and almost fatigue free up to 10$^{10}$ cycles at amplitude of 300 kV/cm being measured. In addition, an enhancement in magnetization, i.e., *Ms*=89.5 emu/cm$^3$, was demonstrated.

## Experiment Procedure

The BFO thin film was deposited by RF magnetron sputtering on Pt-coated Si substrates at 650 °C with LSMO layer of 180 nm in thickness as a buffer layer. To prepare the double layered thin films, polycrystalline ceramic LSMO and BFO targets were prepared from analytically pure oxides by solid state reaction of mixed oxides at 1500 °C for 24 hrs and at 850 for 2hrs, respectively. The detailed experimental procedure for thin film deposition has been reported elsewhere.[13] Phase analyses of the resultant BFO films were made using x-ray diffraction(XRD) (Bruker D8 Advanced XRD, Bruker AXS Inc., Madison, WI, Cu $K\alpha$). Prior to electrical measurements, Au dots of 200 μm in diameter were sputtered on the BFO thin film using a shallow mask to form top electrodes. The cross section and surface morphology of the thin film samples were examined using Field emission scanning electron microscopy (FE-SEM)(Philips, XL30). Their ferroelectric and leakage behaviors were studied by using the Radiant precise workstation (Radiant Technologies, Medina, NY) and a Keithley meter (Keithley 6430, Cleveland, OH). Temperature dependent impedances were obtained using a Solartron impedance analyzer, where the measurement was made from room temperature up to 160 °C in the frequency range of $1–10^5$ Hz. The magnetic behavior of the double layered BFO/LSMO thin films was characterized using a Superconducting quantum interference device (MPMS, XL-5AC, San Diego,CA).

## Results and discussion

Fig. 1 shows the XRD pattern of the BFO thin film buffered with LSMO layer,

where one can see that the BFO/LSMO thin film is of single phase and no secondary phase is detected. All the diffraction peaks are indexed with space group of the *R*3*c*. One can also see that the BFO buffered with LSMO layer gives rise to a highly degree of (111)-oriented film texture, which is similar to what has been observed in BFO/SRO thin film.[9]

Fig. 2(a) shows the surface morphology of the BFO/LSMO thin film by FE-SEM. It exhibits a rather evenly distributed grain sizes together with a dense and smooth surface. The dense texture could greatly contribute to the improved ferroelectric and insulating behaviors.[10] Fig. 2(b) is a cross section image of BFO/LSMO thin film, where the interface is well observed. The cross section also shows a dense polycrystalline texture for both layers, in which the BFO and LSMO layers are 220 nm and 180 nm in thickness respectively, while the Pt/$TiO_2$ layer is 180 nm on Si substrate.

Fig. 3(a) plots the polarization-electric field (*P–E*) hysteresis loops for the BFO/LSMO thin film measured at 3 kHz and room temperature. A well-established hysteresis loop was observed for the BFO buffered with LSMO layer, while the BFO directly deposited on the Pt-coated Si substrate shows a rounded *P-E* loop in association with a high leakage current as shown in inset of Fig. 3(a). It therefore shows that LSMO buffer layer greatly improves the ferroelectric properties of BFO thin film, by dramatically reducing the leakage current.[10] Fig. 3(b) shows the remnant polarization (2$P_r$) and coercive field (2$E_c$), as a function of applied electrical field for BFO/LSMO thin film. At room temperature, a remnant polarization of 2*Pr* ~210.7 μC/cm² was obtained for the BFO/LSMO thin film upon the application of an electric field of 300 kV/cm. The value of $2P_r$ thus obtained is much higher than that of 27 μC/cm² reported previously for a similar system.[14] By comparison, the LSMO buffer layer in the present work gives rise to a highly (111)-oriented texture of BFO, compared to that in the random orientation reported in previous study.[14] Indeed, the $2P_r$ value of 210.7 μC/cm² is comparable to that of BFO/SRO showing (111)-oriented texture.[9] However, the saturated electric field ($2E_c$) measured for BFO/LSMO is 435 kV/cm, which is much smaller than that of 541.1 kV/cm reported in BFO/SRO thin film, implying that LSMO helps to facilitate the domain reversal in BFO.

As show in Fig.3, a well-established *P-E* loop was measured with a much higher $2P_r$ value for the BFO/LSMO thin film, compared to a round like *P-E* loop for the BFO/Pt film. This suggests that the leakage current of the BFO/LSMO thin film is much lower, which can be further confirmed by comparing the leakage currents in the BFO thin films with and without LSMO buffer layers as shown in Fig. 4(a), where the leakage current (*J*) against the applied voltage (*V*) was plotted in semi-logarithmic profiles. One can see that the leakage current in BFO/LSMO thin film is five orders in magnitude lower than that of the BFO/Pt film. There are two likely reasons for the much lowered leakage current including: (i) the interfacial diffusion between BFO and Pt, which is a common phenomenon for ferroelectric thin film deposited on Pt electrode, was depressed by the LSMO buffer layer;[10] (ii) the dense film texture and uniform grain size can also play a role in the reduced leakage current.

For BFO thin films, the electric behavior is largely influenced by various defects present. We have therefore employed the complex impedance to identify the likely types of charge carriers. The frequency-dependent imaginary part of modulus ($M''$) for the BFO/LSMO thin film is shown in Fig. 4(b). A single relaxation peak is observed in the film at each temperature and it shifts towards the higher frequency side with increasing temperature, indicating the occurrence of a thermally-activated process. To determine the nature of defects involved in the BFO/LSMO thin film, the activation energy can be obtained using the Arrhenius law:

$$f_m = f_0 \exp(-\frac{Ea}{K_B T}) \qquad (1)$$

where $f_m$ is the hopping frequency, $f_0$ is the pre-exponential factor, and $Ea$ is the activation energy. As shown in the fitting results plotted in inset of Fig. 4(b), the activation energy thus obtained is 0.89 eV, which is comparable to 0.99 eV measured for SrTiO$_3$,[15] indicating the oxygen vacancies are the predominant defects existing in the BFO/LSMO thin film. Indeed, oxygen loss can further be accelerated and lead to the formation of more oxygen vacancies, during the sputtering process at high temperatures (e.g., 650 °C in the present work) and low oxygen partial pressure ($10^{-2}$ mbar). The process can be expressed as:

$$O_O^x \rightarrow V_O + \frac{1}{2}O_2 \qquad (2)$$

The calculated value of activation energy in this work suggests that the oxygen vacancies are largely responsible for the conduction in BFO/LSMO thin film.

Fig. 4(c) shows the relative dielectric constant ($\varepsilon_r$) and dielectric loss (tanh δ) as a function of frequency for the BFO thin film with and without LSMO buffer layer. It shows that the BFO/LSMO thin film exhibits a much higher dielectric constant and lower loss (i.e. 116.8 and 2.7% at 1 kHz) than those (i.e 59.8 and 18% at 1 kHz) of the BFO/Pt film, in the frequency range from 1 to $10^6$ Hz investigated in the present work. The BFO/Pt thin film is apparently dominated by the space charges, as shown in Fig. 4(a), as a result of the secondary phase and occurrence of structure defects,[10] which are also responsible for the poor dielectric properties. In contrast, the BFO buffered with LSMO exhibits much a dense columnar film texture and a lower leakage current, which can contribute towards the enhanced dielectric properties.

Fig. 4(d) shows the ferroelectric fatigue endurance as a function of polarization switching cycles for the BFO/LSMO thin film measured at the frequency of 500 kHz with amplitude of 300 kV/cm. Compared to the serious fatigue failure observed for the BFO/Pt film,[16] a significant improvement in fatigue resistance was observed for the BFO buffered by LSMO layer, where it is almost fatigue-free up to $10^{10}$ cycles. For the BFO/Pt thin film, interfacial diffusion between BFO thin film and Pt-coated Si substrate is expected to take place and these space charges (i.e oxygen vacancies) play an important role. The piling up of oxygen vacancies at the interface can apparently pin domain walls, thus leading to an apparent fatigue failure.[17] The presence of LSMO as an oxide conductive buffer layer can act as a sink for oxygen vacancies,[18] resulting in a reduction in the concentration of oxygen vacancies. Consequently, the fatigue endurance can be dramatically improved.

The magnetic behaviors measured for BFO/LSMO double layers and LSMO single layer at room temperature are illustrated in Fig. 5(a). In order to illustrate the contribution of the LSMO buffer layer towards the magnetization, the magnetization values of BFO/LSMO double layers and LSMO single layer are both normalized with areas instead of volumes. Saturated loops were observed for both BFO/LSMO and LSMO thin films and the normalized magnetization values are comparable to each other. It therefore shows that the greatly enhanced magnetization occurring in the BFO/LSMO double layers is largely contributed from the LSMO buffer layer, in which the magnetization arises from the double exchange interaction between $Mn^{3+}$ and $Mn^{4+}$.[19] We have further observed that the magnetization value obtained for the BFO/LSMO thin film is slightly higher than that in the LSMO single layer. This much enhanced magnetization may come from a diffuse interface boundary formed by the diffusion of lanthanum and manganese from LSMO into BFO.[14] Fig. 5(b) shows the temperature-dependent magnetic hysteresis loop of the BFO/LSMO thin film with the field applied in the direction parallel to the film surface. In the temperature range from 10 to 300 K, a well-behaved *M-H* curve was observed for the BFO/LSMO double layers and the magnetization increases with increasing applied field before reaching the saturated value, as indicative of a typical ferromagnetic behavior. The magnetization is temperature dependent above 100 K, where the saturated magnetization is dramatically enhanced with a decrease in temperature. In contrast, the saturated magnetization is almost temperature-independent and keeps an almost constant value below 100 K. The in-plane coercive fields measured are 26, 108, and 158 Oe at 300, 100, and 10 K, respectively, indicating that there is a close relationship between the coercive field and temperature. Therefore, the coexistence of ferroelectric and ferromagnetic orders occurs in the double layered BFO/LSMO thin film, suggesting that it may be a promising candidate system for the pursuit of the magnetoelectric effect.

## Conclusions

Multiferroic BFO thin films were successfully deposited on LSMO buffered Pt/TiO$_2$/SiO$_2$/Si substrates by off-axis RF magnetic sputtering, where a high degree of (111)-oriented texture was observed. The texture development and electrical behaviors of BFO thin film were greatly improved by employing LSMO as the buffer layer. A remnant polarization of 210.7 μC/cm$^2$ in 2$P_r$ and a coercive field of 435 kV/cm in 2$E_c$ are measured for the BFO thin film buffered with LSMO. The thin film exhibits a relative permittivity of 116.8 and a dielectric loss of 2.7% at 1 kHz. It is almost fatigue-free when switched at 500 kHz up to 10$^{10}$ switching cycles, attributed to the highly desirable film texture arising from the LSMO buffer layer. The thin film also shows a notable improvement in magnetic behavior ($Ms$=89.5 emu/cm$^3$), which is largely resulted from the presence of the LSMO buffer layer.

Captions for Figures:

FIG. 1. (Color online) XRD patterns of the double layered BFO/LSMO thin film measured at room temperature.

FIG. 2. (Color online) (a) SEM image showing the surface morphology, and (b) typical cross-section of the BFO/LSMO thin film deposited on the Pt coated Si substrate.

FIG. 3. (Color online) (a) *P-E* loops of the BFO thin film with LSMO buffer layer, and the inset shows the *P-E* loop for the BFO/Pt thin film. (b) $2P_r$ and $2E_c$ as a function of the electric field.

FIG. 4. (Color online) (a) Leakage current as a function of the applied voltage for BFO thin films with and without LSMO buffer layer, (b) Variation of $M''$ as a function of frequency, where the inset shows the Arrhenius plot of the hopping frequency against temperature, (c) Frequency dependence of dielectric properties for BFO thin film with and without LSMO layer, and (d) Fatigue endurance of the BFO thin film with LSMO buffer layer.

FIG. 5. (Color online) (a) Normalized magnetic moment by areas as a function of the applied field for LSMO single layer and BFO/LSMO double layers at room temperature. (b) *M-H* loop for BFO/LSMO at varied temperatures.

**FIG. 1.**

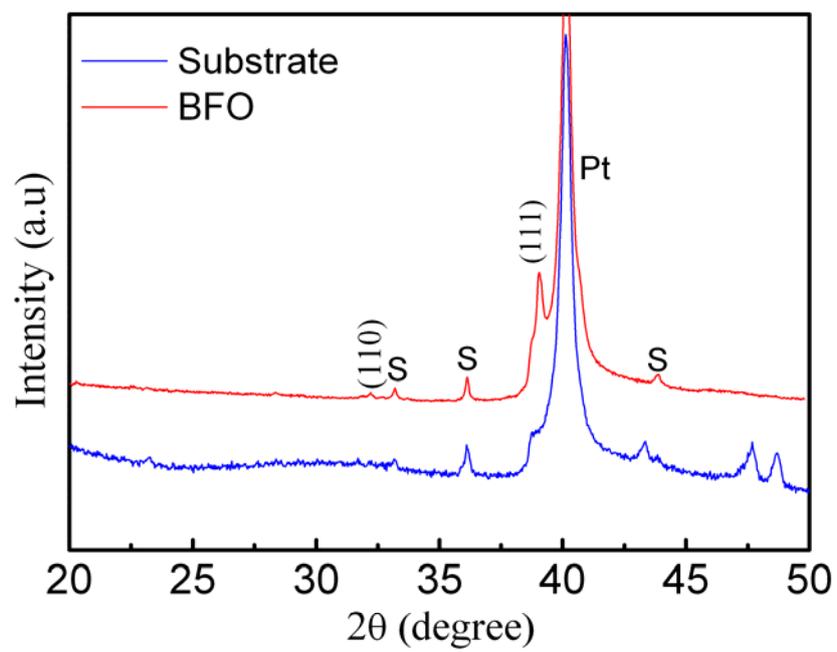

**FIG. 2.**

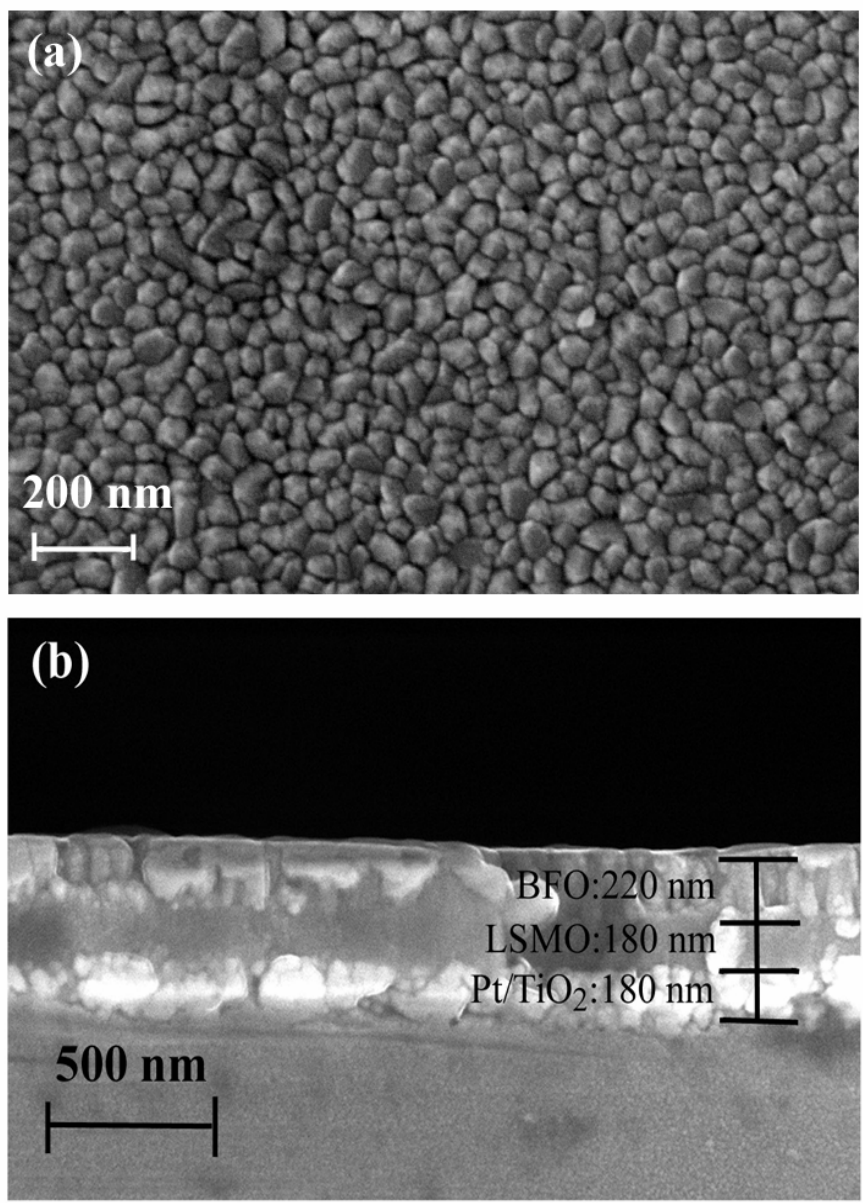

**FIG. 3.**

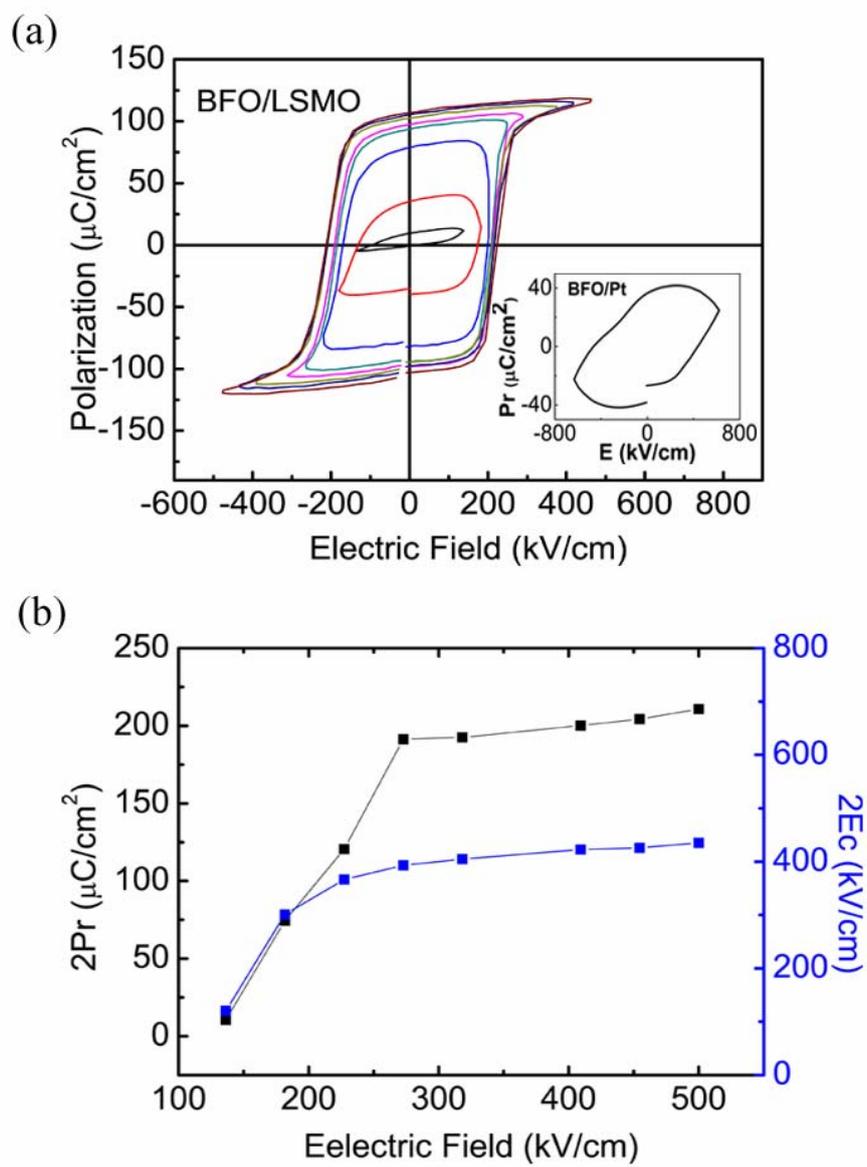

**FIG. 4**

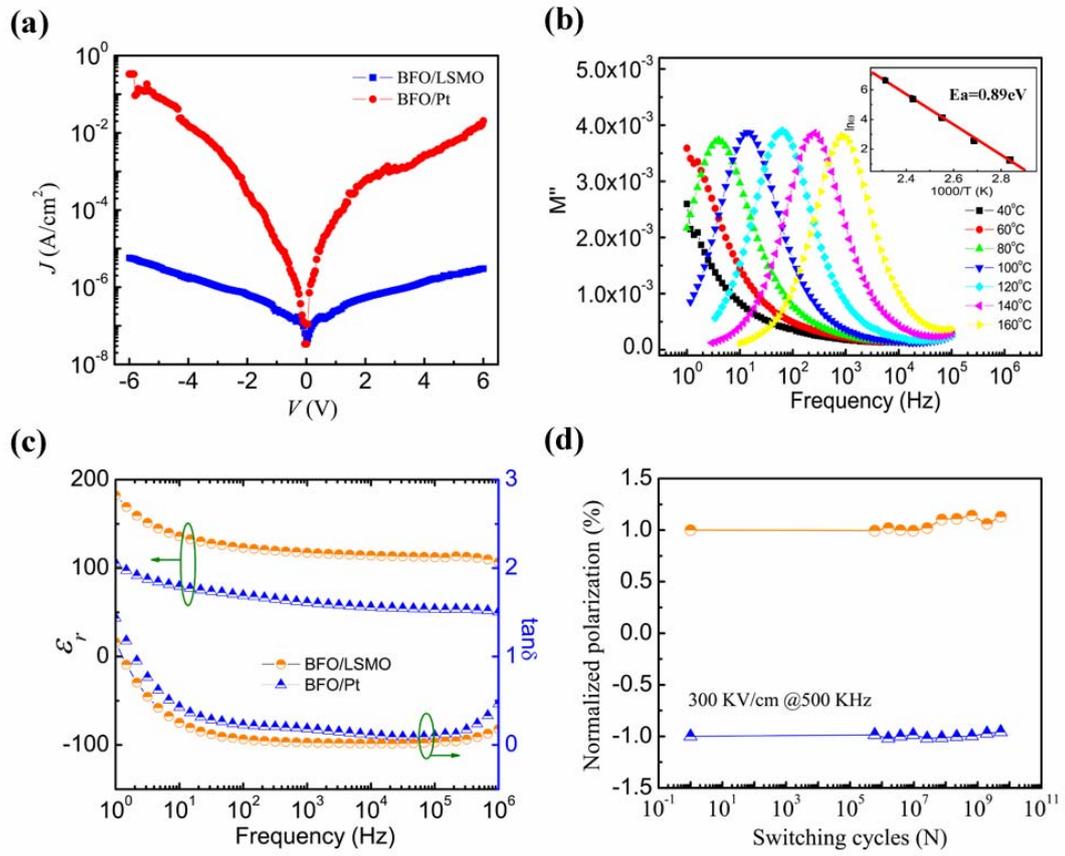

**FIG. 5**

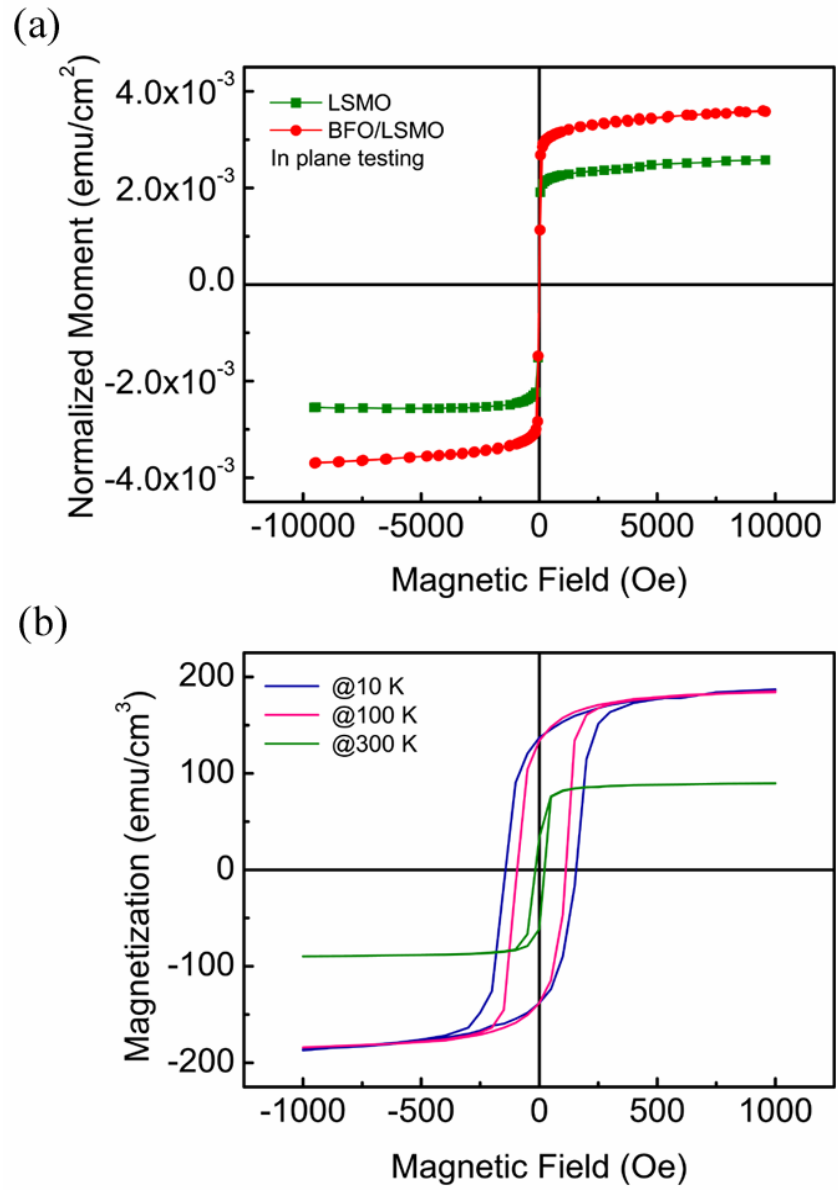